\documentclass[runningheads]{llncs}
\usepackage[T1]{fontenc}
\usepackage{graphicx}

\usepackage{fontawesome5}

\usepackage{colortbl}
\usepackage{xcolor}

\usepackage{subfig}
\usepackage{float}

\usepackage{tabularx}
\usepackage{booktabs}

\usepackage{cite}

\begin{document}


\title{``I Feel Like I'm Teaching in a Gladiator Ring'': Barriers and Benefits of Live Coding in Classroom Settings}

\author{Caroline Berger\inst{1} \and David Weintrop\inst{2} \and Niklas Elmqvist\inst{1}} 

\institute{Aarhus University, Department of Computer Science, Aarhus, Denmark \email{\{caroline.berger\}\{elm\}@cs.au.dk} \and
University of Maryland, College Park, Maryland, USA
\email{weintrop@umd.edu}}

\authorrunning{C. Berger et al.}

\maketitle

\begin{figure}[H]
\includegraphics[width=\textwidth]{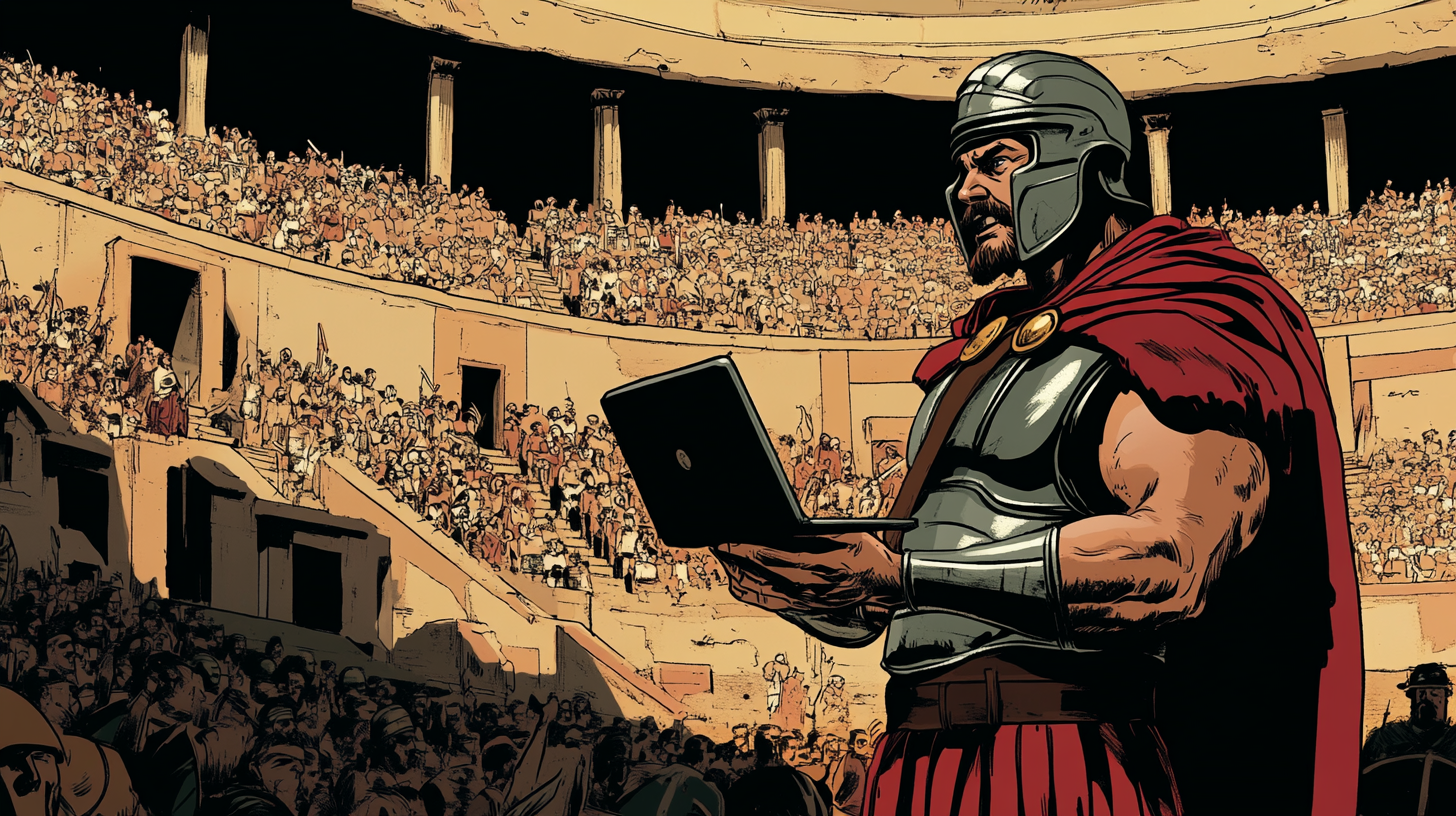}

  \caption{\textbf{Live coding as a performance.}
  Our study found that many instructors feel live coding to be a high-stakes and cognitively taxing activity more akin to a live performance.
  (Illustration by MidJourney version 6.1.)
  }
  \label{fig:teaser}
\end{figure}

\begin{abstract}
    Live coding for teaching---synchronously writing software in front of students---can be an effective method for engaging students and instilling practical programming skills.
    However, not all settings are conducive to live coding and not all instructors are successful in this challenging task. 
    We present results from a study involving university instructors, teaching assistants, and students identifying both barriers and benefits of live coding. 
    Physical infrastructure, a positive classroom community with psychological safety, and opportunities for teacher development are practical considerations for live coding.
    In order for live coding to be an active learning experience, we recommend that tools support multiple mechanisms for engaging students, directing audience attention, and encouraging student-led live coding.
\end{abstract}



\section{Introduction}

\begin{quote}
    \textit{``And then I'm teaching [...] in a classroom that feels like a gladiatorial ring. 
    200 seats in a wall up in front of me. 
    And I have to lean back to see the top. 
    And really the only constraint in that classroom is that it's terrifying. 
    It is the most terrifying experience I've ever had.''}
    -- Participant 08 (Computer Science instructor)
\end{quote}

The gladiator descends into the arena engulfed by the shrill screams of the spectators.
Rows upon rows of the audience stares down at him, anticipating the long-awaited battle.
But we aren't in Ancient Rome; we are in a giant tiered lecture hall on a present-day university campus.
The spectators are students learning programming.
And the gladiator? 
It's the instructor, equipped with only their laptop and some notes, and the feat that they are about to attempt is to write valid source code in front of an auditorium full of students.
While an effective means of engaging the audience and conveying authentic, practical programming skills, instructors liken this form of \textit{live coding} to a performance or---as P08 puts it above---a gladiatorial battle because of its high stakes, captive audience, and technical challenge (Figure~\ref{fig:teaser}).

Live coding is defined as \textit{``the process of writing code live on a computer in front of students during class''}~\cite[p. 164]{DBLP:conf/iticse/SelvarajZPR21}.
Live coding models the process of programming~\cite{ali2005effective} and gives students insight into practical programming skills and practices. 
However, this performative form of instruction does not always result in active engagement from the students, which is an important component of effective learning~\cite{michael2006s}.
Fostering engagement requires instructor preparation and time~\cite{DBLP:conf/sigcse/CaceffoGA18}, and risks increasing workload and burnout~\cite{el2016teachers}
Additional work is needed to gather the perspectives and experiences of instructors on active learning during live coding activities~\cite{DBLP:conf/iticse/SelvarajZPR21}. 
In addition to instructor-led live coding, student-led live coding (i.e., a student coding in front of their peers) as a form of peer instruction requires further investigation~\cite{DBLP:conf/iticse/SelvarajZPR21}.
The perspectives of students and instructors can provide essential insights into design considerations for creating effective and easily adopted live coding tools. 
It is this gap in the literature that this work seeks to address.

Motivated by this gap in the knowledge for live-coding challenges and design considerations, our work pursued the following research questions (RQ1-3):
(1) What are the perceived barriers to live coding?
(2) What are the perceived benefits of live coding? and
(3) What are best practices and technological considerations to support this practice in classrooms?
To answer these questions, we conducted a study involving a range of live coding stakeholders, including the instructors and teaching assistants who perform these sessions, and the students who participate in them.
We conducted semi-structured interviews with single participants as well as groups, using prototypes as springboards.
Our thematic analysis of these sessions resulted in the identification of perceived \textbf{barriers} that make effective live coding challenging, as well as \textbf{benefits}  associated with the approach.

In summary, we found that stage fright, high cognitive load, and poor prior experiences make live coding difficult.
Due to the negative emotions associated with live coding, instructors often avoided asking students to live code in front of their peers, thus missing an opportunity for peer instruction and active learning.
In particular, instructors and students reported the emergence of intentional and unintentional mistakes and the ability to involve the class as benefits of live coding. 
Our analysis also revealed how personal computers might detract from live coding lectures and challenges associated with directing student attention to important parts of the instructor's code. 
Based on our findings, we recommend a learning environment that encourages risk taking, as well as opportunities for professional development through mentorship and a shift of mindset for teachers.
To support student participation, we recommend tools be designed to increase comfort and reduce performance anxiety for students coding in front of the class.

\section{Related Work}

This work is situated at the intersection of  education, computer science, and human-computer interaction. 
Here we examine the current state of live coding tools and practices.

\subsection{Active Learning in Programming Instruction}
A significant body of research has investigated the challenges associated with teaching novices how to program~\cite{luxton2018introductory, robins201912}.
This research has attended to motivational and sociocultural aspects~\cite{malmi2020theories}, pedagogy~\cite{falkner201915}, the cognitive complexity of the topic~\cite{qian2017students}, and the design of programming languages and environments~\cite{mcgill2020construction}. 
Across the insights from this work, active learning has emerged as a productive instructional practice with a variety of instantiations in contemporary computer science classrooms~\cite{berssanette2021active}. 
As opposed to \textit{passive learning}, where the instructor conveys information with minimal student involvement, \textit{active learning} entails continuous interaction between students and instructor, supporting the student's engagement in the process of learning~\cite{bonwell1991active}. 
Despite active learning having shown to be effective for STEM (Science, Technology, Engineering, and Mathematics) subjects~\cite{freeman2014active}, passive instruction is still common in computer programming education~\cite{DBLP:journals/te/FroydBCHP13, DBLP:journals/jeric/GrissomMM17}.

Learning to program requires practice with feedback to help the student guide their own learning.
This kind of deliberate practice---practice that is coupled with continuous feedback---is vital for acquiring the necessary cognitive, perceptual, and motor skills~\cite{ericsson1993role}.
Such feedback might come from programming tools as error messages or syntax highlighting, or in classroom settings from instructors, teaching assistants, or peers. 
In particular, the feedback can be delivered at a scale to a classroom of students through \textit{live coding}: writing code live on a computer in front of a class~\cite{DBLP:conf/iticse/SelvarajZPR21}.

Several active learning strategies have been developed and studied in computer science contexts, including flipping the classroom, project-based learning, Process Oriented Guided Inquiry Learning (POGIL), and peer instruction ~\cite{berssanette2021active, DBLP:conf/sigcse/PorterBCGLMZS16, crouch2001peer, POGIL_Yadav_2019}.  
Students appreciate active learning in computer science, with reports of increased satisfaction, motivation, and confidence, along with improved learning outcomes over traditional techniques~\cite{berssanette2021active}.
Centering the student in their learning, active learning techniques help students to learn at their own pace. 
Despite positive reactions and improved educational outcomes of active learning, such techniques require additional effort for instructors when compared to traditional techniques~\cite{berssanette2021active}. 
In this paper, we aim to examine possibilities for active learning in the context of live coding in ways that minimize increased instructor workload.

\subsection{Live Coding Practices}

A common technical setup for live coding is for the instructor to simply project their integrated development environment (or share their screen) while they are live coding~\cite{DBLP:journals/ploscb/NederbragtHHW20}.
Students can then follow along in their own editors. 
The instructor often talks aloud to provide rationale for the actions taken~\cite{DBLP:journals/ploscb/NederbragtHHW20, DBLP:conf/iticse/SelvarajZPR21} and encourages the students to ask questions during the process~\cite{DBLP:journals/ploscb/NederbragtHHW20}. 
Some students prefer to code along with the instructor~\cite{DBLP:conf/kolicalling/RajPHH18}, while others do not~\cite{lauvaasexperience}. 

A flipped classroom approach where students watch a prerecorded video of an instructor writing code is another flavor of live coding that has shown to be effective for learning programming~\cite{lin2022teaching}.
Associated with positive learning outcomes, the live-coding videos for learning programming (LV4LP) platform supports self-regulation by providing the ability for students to pause and rewatch portions of the video, and opportunities for students to reflect by annotating timestamps of the video~\cite{lin2022teaching}. 
However, such approaches are not strictly speaking ``live'' in that they do not allow the instructor to interact directly with the students and respond to their questions in real time.

While there remain open questions as to the specifics of its benefits and limitations relative to other teaching methods, live coding is a promising direction due to positive reactions from instructors and students~\cite{DBLP:conf/sigcse/Rubin13, DBLP:conf/iticse/SelvarajZPR21, shah2024comparison}.
Live coding lessons in which students code along for a portion of the lesson, incorporating active learning, lead to more engagement then traditional instructor only live coding~\cite{shah2024comparison}.
Research has found live coding to be helpful in reducing the extraneous cognitive load on students~\cite{Raj_el_al_2020_LiveCodingCogLoad} and to encourage more question asking and participation~\cite{watkins2024comparing} when compared to comparable static code learning experiences.
It is also a scalable delivery method because of its one-to-many relationship between instructor and students.
Importantly, live coding models the incremental process of programming~\cite{ali2005effective} rather than instantly displaying a finished product that, once run, will produce an error-free output.
In contrast, live coding presents a more realistic picture of the practice of programming which can be important for easing novice students into the craft and complexities of programming.

\subsection{Live Coding Tools}

Simultaneously writing correct code, describing the code you are writing and managing the technology used for sharing the source code with the students can be a significant challenge for instructors.
Several tools have been developed to help instructors manage the live coding challenge. 
For example, Improv~\cite{DBLP:conf/lats/ChenG19} was developed to support live coding by synchronizing blocks of code with presentation slides and allowing instructors to set waypoints to link the progress in the live coding example with conventional instructional slides.
To involve students in coding exercises, VizProg~\cite{zhang2023vizprog}, helps instructors monitor live learning by displaying student progress towards a solution, while Overcode~\cite{DBLP:journals/tochi/GlassmanSSGM15} analyzes students’ submissions.
Another tool for in class coding---Codeopticon~\cite{DBLP:conf/uist/Guo15}---gives a gallery view of students' code so that instructors can track progress and identify potential discussion points based on student work.

A core strength of live coding is the ability to react to the classroom and adjust materials in real time~\cite{guzdial2021teaching}. 
However, existing tools such as VizProg ~\cite{zhang2023vizprog} and Overcode ~\cite{DBLP:journals/tochi/GlassmanSSGM15} are designed to support learning in progress towards a correct solution, which implies the requirement of exercise test cases to be prepared and loaded into the system before the lesson, thus increasing instructional workload. 
In addition to adding preparation requirements on the instructional team, by predefining the exercises, the instructor is not able to create activities that respond to the in-class environment, the progress of the lesson, student’s questions, and difficulties~\cite{nederbragt2020ten}. 

An alternative approach is to provide feedback to the instructor (and students) about trial-and-error, but not necessarily correctness.
This gives instructors greater liberty to create exercises during class time and allows for student expression in the results that might not be possible by predefining output, for example, by connecting the answer to an important part of their own lives.

\section{Method}

In pursuit of answering the stated research questions, we performed a study with instructors, teaching assistants, and students with direct experiences in teaching and learning programming via live coding in large lecture halls and teaching assistant sessions. 
The recruitment materials, demographics questionnaire, the facilitator guide that includes the interview questions, participant demographics, positionality statement, and coding materials are available at \url{https://osf.io/rq8bd/?view\_only=420c9f015e804297bf7b3d3838e3d317}. 

\subsection{Participants}

We recruited participants through the personal Linkedin pages of the research team and university mailing list servers.
Two instructors, seven teaching assistants, and six students participated in the study. 
All participants were 18 years or older, spoke English, and had either taught or attended a post-secondary course that included programming in the past year. 
Demographic information about participants is displayed in Table~\ref{tab:firsthalfparticipants}.
With the exception of two participants (P02 and P03), the study was conducted with one participant and the first author. 
P02 and P03 expressed a preference to participate in the study together, so the first author conducted the study with the two participants at the same time. 
Although P02 is a teaching assistant, they have not taught P03 (a student).

Thirteen participants identified as men, one participant identified as a woman, and one participant identified as non-binary.
Although we did not ask, one participant disclosed that they had ADHD and another disclosed that they had ADHD, dyslexia, and autism. 
Including the perspectives of neurodiverse individuals added a layer of richness to the findings.

\begin{table*}
    \centering
    \begin{tabular}{cllcll}
    \toprule
    \textsc{\#} & \textsc{Role} & \textsc{Setting} & \textsc{Age} & \textsc{Gender} & \textsc{Education} \\
    \midrule
    
    \rowcolor{gray!10}
    P01 & Teaching Assistant & Online & 26-35 & Man & Ph.D. Info. Sci. student \\   
    P02 & Teaching Assistant & In-Person & 18-25 & Non-Binary & B.Sc. CS student \\
    \rowcolor{gray!10}
    P03 & Student & In-Person & 18-25 & Man & B.Sc. CS, Math student\\
   P04 & Student & In-Person & 18-25 & Man & M.Sc. HCI student \\
   \rowcolor{gray!10}
   P05	& Student & In-Person & 26-35 & Man	& M.Sc. HCI student\\
   P06 & Teaching Assistant & In-Person & 18-25 & Man & B.Sc. CS student \\ 
   \rowcolor{gray!10}
   P07 & Teaching Assistant & In-Person & 26-35 & Man & Ph.D. HCI student\\
   P08	& Instructor & In-Person & 36-45 & Man & M.Sc. Info. Sys.\\
   \rowcolor{gray!10}
   P09 & Student & In-Person & 18-25& Man & B.Sc. CS student\\
   P10 & Student & In-Person & 18-25 & Man & M.Sc. HCI student\\
   \rowcolor{gray!10}
   P11 & Instructor & In-Person & 45+ & Man & Ph.D. Info. Sci. student\\
   P12 & Teaching Assistant & In-Person &	18-25 & Man & B.Sc. CS student \\
   \rowcolor{gray!10}
   P13 & Student & In-Person & 18-25 & Man & B.Sc. CS, Robotics student\\
   P14 & Teaching Assistant & Online & 18-25 & Woman & Ph.D. CS, CS Ed. student\\
   \rowcolor{gray!10}
   P15 & Teaching Assistant & In-Person & 26-35 & Man & Ph.D. Info. Sci. student\\
   \bottomrule
   \end{tabular}
    \caption{\textbf{Participant demographics.} Participant roles, study setting, age, gender, and education background.}
    \label{tab:firsthalfparticipants}
\end{table*}

\subsection{Prototype}
\begin{figure}
    \centering
    \includegraphics[width=\linewidth]{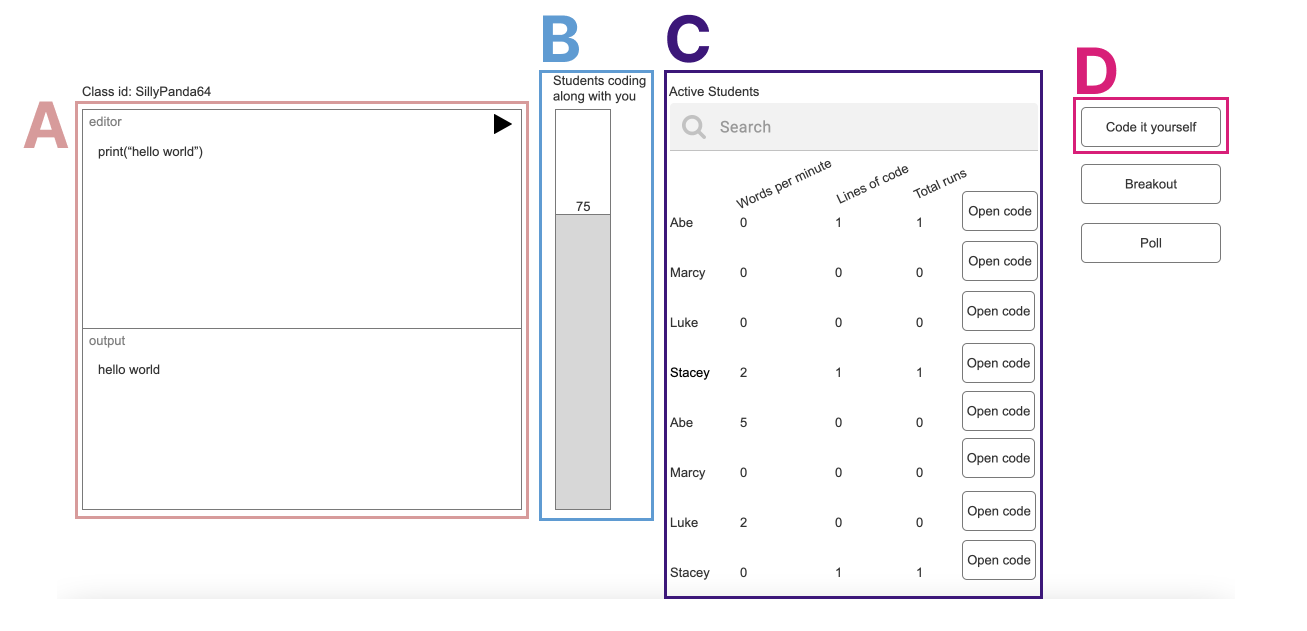}
    \caption{{\textbf{Teacher live coding view.}
    The teacher can code in their editor (A). 
    The number of students coding along with the instructor is displayed in the meter (B) and information about students is in the active student pane (C). 
    The teacher can start an independent activity by selecting code it yourself (D).}}
    \label{fig:TeacherMainView}
\end{figure}
\begin{figure}[htb]
    \centering
    \includegraphics[width=1\textwidth]{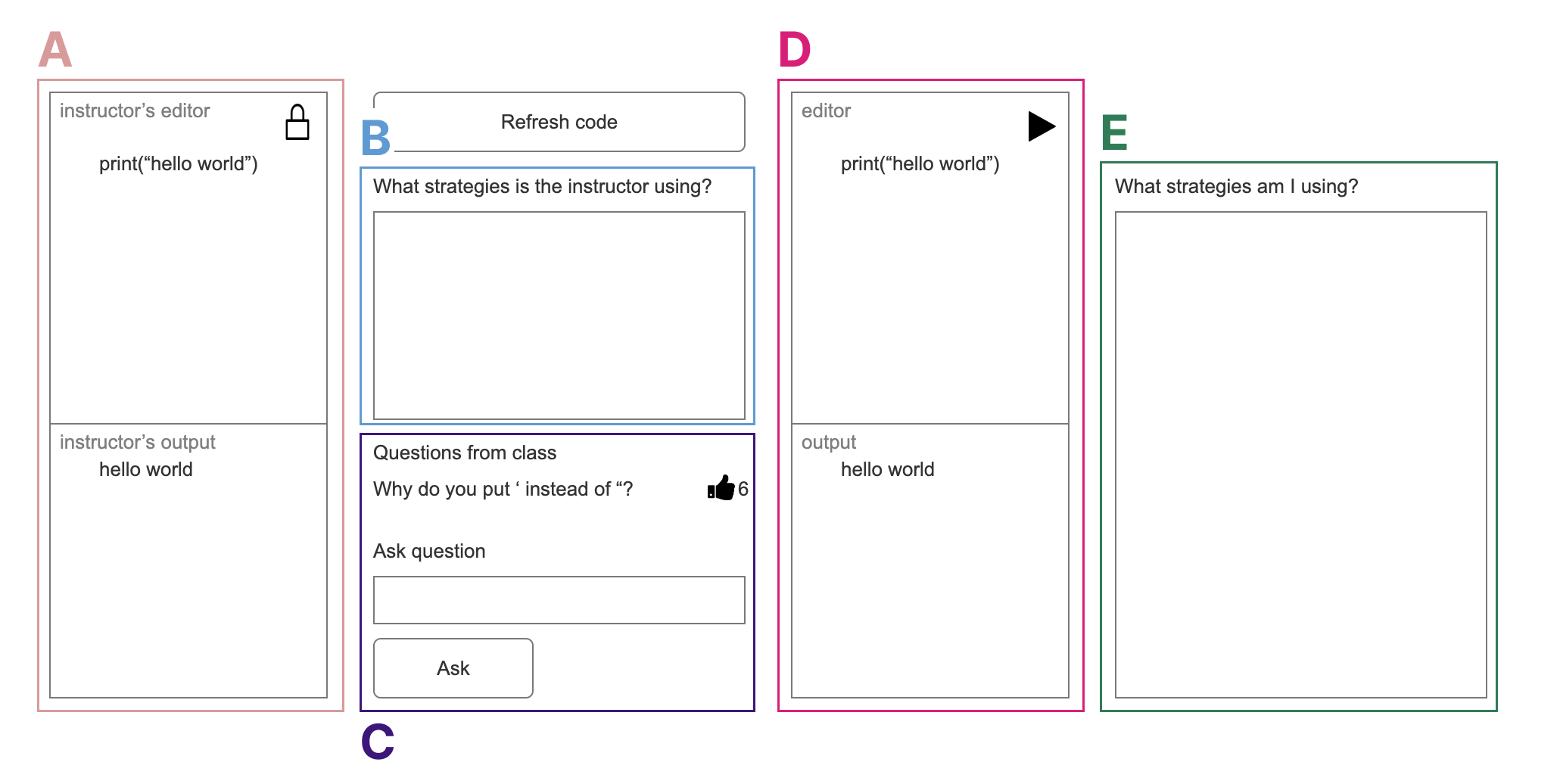}
    \caption{\textbf{Student coding along with the instructor prototype view.} 
    Student view has the instructor's editor (A), their own editor (D), and notepads (B, E) where the student articulates the strategies they use and the instructor uses while coding.
    Refresh code updates the students code to match the instructors code. 
    The student can ask questions (C).}
    \label{fig:studentcodingalong}
\end{figure}
Our prototype live coding tool (available at \url{https://gc9vfa.axshare.com/}) was designed to encourage class participation in live coding. 
To encourage divergent brainstorming, the aesthetics and the interactions of the prototype are low fidelity: black and white colors and primal buttons over features that invite complex interaction. 
It asks students to articulate the strategies the instructor uses and then employ them themselves as they work through the programming activities (Figure~\ref{fig:studentcodingalong}). 
We build on prior work that explored reflection of programming strategies with professional software engineers~\cite{DBLP:conf/chi/ArabLLK22}.
The prototype is informed by cognitive apprenticeship~\cite{collins1991cognitive, collins2014cognitive}, encouraging lessons to go beyond modeling to reflect on programming strategies and to engage the student in experimenting with code alongside the instructor (Figure~\ref{fig:TeacherMainView} and Figure~\ref{fig:studentcodingalong}).
Prior work recommends coding along during live coding as an avenue for continued exploration~\cite{DBLP:conf/ace/HwangALRSR21, shah2024comparison}.

\subsection{Procedure}

Participants were asked to complete a demographics questionnaire (age, gender, and educational background) before participating in the study. 
Inspired by contextual inquiry techniques~\cite{karen2017contextual}, the sessions aimed to capture accounts of participants' lived experience. 
In general, our study is based on a research-through-design methodological approach~\cite{DBLP:books/sp/14/ZimmermanF14}.
The approach relies on prototypes that ground communication and serve to externalize ideas and hypothesis, and invite critique and continued ideation. 
Prototypes, in their physicality, serve as a sounding board for ideas to spring. 

The sequencing is loosely inspired by the future workshops method~\cite{vidal2006future} as participants critique current live coding practices in the first interview and fantasize about live coding in a perfect world without constraints in the closing interview. By reviewing the prototype between critique and fantasy, the participants interact with a possible design intervention to ignite ideation of other possible solutions.

The studies occurred in a U.S.\ public university setting: two on Zoom and thirteen in-person. 
We recorded participant's voices. 
The study duration was between 30 minutes and 1 hour and occurred in the spring semester in late February and early March of 2023.
All participants gave their informed consent before participating in any of the study activities, and the study was approved by our university's institutional review board. 

\subsection{Positionality}
As the methods and analysis are shaped by the researchers and their prior experiences, we discuss the lens the research team approaches the work.
Demographics-wise, the author who facilitated the study and performed the thematic analysis is a white, American, neurotypical cis woman in her twenties. 
She started programming during high school in electives and at an afterschool programming club and studied computer science in her undergraduate degree at a large public university in North America. 
She learned to program largely through attending lectures where professors would live code and enjoyed this experience. 

During the summers, she taught programming and robotics where she live coded often and asked students to live code in front of their peers. 
She observed moments of pride when students demonstrated mastery in front of their classmates. 
She holds the belief that anyone can learn to program given resources, time, and motivation. 
She is skeptical if technology and new tools are the answers to the problems faced in computer science education or if technology is the source and contributing force to the problems.
Unimpressed with the gender and racial diversity in her undergraduate classes, she strives to make computer science a more welcoming space.

\subsection{Data Analysis}
We analyzed the data through thematic analysis~\cite{clarke2015thematic} guided by practices in analyzing interview data~\cite{weiss1995learning} with attention to types and richness of information over counts and frequency~\cite{creswell2016qualitative}. 
In a reflexive manner, codes emerged from the data with a focus on the ways in which students and instructors reflect on their live coding experiences. 
Descriptions, barriers, benefits, and design opportunities of live coding were codes that emerged from the data. 
Table~\ref{tab:codingofsignificantstatements} presents the code, associated note, and an example quote for the emergent themes that were identified.

\begin{table*}
   \centering
   \begin{tabular}{llp{0.47\linewidth}}
   \toprule
   \textsc{Code} & \textsc{Note} & \textsc{Quote} \\
   \midrule
   \rowcolor{gray!10}
   Description & Live coding as a performance & \textit{``It's like live performance.
   It's really hard to practice it enough that you know that it's going to work, but also have that kind of ability to take student suggestions and potentially go in a direction that you haven't tested and might not work out.''} (P08)\\
   
   Barrier	& Fear of messing up &
   \textit{``Part of it is the pressure of just being in front of an audience.
   And you sort of, I mean, naturally you don't want to mess up.
   And so thinking of that gives you some sort of, I guess, anxiety, but I guess for me over time at first I was definitely like nervous since it was my first time doing anything like that.
   But I think in my experience, I got less nervous and much more comfortable.
   But yeah, I think the main thing is definitely just the anxiety of messing up so badly for students.''} (P06)\\  
   \rowcolor{gray!10}
   Benefit & On-the-fly nature of live coding &
   \textit{``I'm doing some example, then it's easier to change stuff on the fly and then surprise students.''} (P02)\\
  Design opportunity & Gallery camera view &
  \textit{``Okay, these students got it these students didn't.
  I would love to have a second screen that had, you know, the small kind of security camera view where I had every student desktop and be able to see that they're all on their own.''} (P11)\\
  \bottomrule
  \end{tabular}
  \caption{\textbf{Excerpts of coding process.}
  Interviews were coded into descriptions, barriers, benefits, and design opportunities with accompanying notes.}
\label{tab:codingofsignificantstatements}
\end{table*}

After identifying codes, we went through the interviews to identify excerpts that described the codes. 
To integrate data, we created clusters.
We analyzed significant statements to generate themes~\cite{creswell2016qualitative}.
After identifying themes, we went through interviews to find additional evidence to support the themes.
A colleague reviewed the preliminary groupings of data and gave feedback to improve the clarity of our resulting themes.

\section{Findings} 

By interviewing instructors, teaching assistants, and students, we uncovered barriers and benefits of live coding. 
Themes arose from the participant responses within the categories of barriers, benefits, and design opportunities. 
Below we present each theme, using participant’s own words to illustrate their central characteristics.

\subsection{Barriers of Live Coding}

\paragraph{Teaching environment: inadequate resources and setup}

The first barrier our analysis revealed is related to the larger environment in which live coding takes place. 
Attributes of the classroom and the lack of necessary resources make live coding difficult. 
In general, \textbf{organizing the physical lecture space} is difficult during live coding lectures.
This challenge is compounded when instructors are also keeping a correct copy of the code open for reference---often as a physical printout---as it introduces yet another thing to be kept visible and managed in a limited space.

\begin{quote}
    \textit{... because a lot of the classrooms I'm struggling to find laptop space, space for my printouts... that are convenient, you know, lecturing is one thing, having the paper accessible as I'm typing is another challenge.} (P11, instructor)
\end{quote}

Indeed, unlike slide-only lectures, during live coding there are several pieces of information that the instructor must juggle at once. 
For example, the instructor manages content slides and their live coding environment on a computer. 
One instructor identified \textbf{managing lecture slides and the editor} as challenging, which corroborates prior work~\cite{DBLP:conf/lats/ChenG19}:

\begin{quote}
    \textit{I mean, I also struggle just with logistics of switching back and forth between the PowerPoint [slides] that I'm running and terminal to do coding in...} (P11, instructor)
\end{quote}

From the student perspective, the jump between slides and coding environments can create confusion.

\begin{quote}
    \textit{...[the instructors] explain [loops] within a slide deck... this is how loops work...
    But then they would give an example and switch over to whatever coding environment and then [the instructors] show that. 
    And there's a separation of those two things, which is a little confusing.}
    (P10, student)
\end{quote}

Another challenges stems from the limitations of the projector, which is essential during live coding as it displays the code for students to see.
Text must be large enough that students sitting at the back in hundred person classrooms can see the code, yet not too large that the code is unmanageable in the instructor's editor.
Said P10, a student who regularly attended live-coding lectures:

\begin{quote}
    \textit{I think it would have been nice to have some way to follow along better with live coding, [...]
    especially because my lectures were like 150, 200 people and [it's] such a big class and sometimes you sit in the back and...
    [...]
    I realized quite soon that I have to sit all the way in the front for me to pay attention in class, especially in those big classes, because if you sit in the back you get distracted, but also you can't really see what [the instructor is] doing.} (P10, student)
    \label{distractorquote}
\end{quote}

An instructor reported struggling to get correct \textbf{projector to screen ratio} so that both student and instructor can see the code. 

\begin{quote}
    \textit{[The room has] monster screens and there's four of them, but if you don't have eagle eyes, you can't see them. 
    It's still too small. 
    So I have Visual Studio blown up to the max so that they can see it. 
    So what is the problem there? 
    It's the amount of different windows that there is and the text isn't big enough. 
    It's just trying to get the text big enough, but also at a point that I can use it on my screen.} (P08, instructor)
\end{quote}

\paragraph{Live coding is scary}

By interviewing instructors, teaching assistants, and students, we found that live coding can be an emotionally fraught experience.
Participants reported feelings of \textbf{nervousness} and having \textbf{stage fright}.
Since a common live coding setup is one person in front of many, the attention is focused on the performer.
Furthermore, the classroom layout can create feelings of anxiety while the instructor live codes. 

\begin{quote}
    \textit{And then I'm teaching [...] in a classroom that feels like a gladiatorial ring. 
    200 seats in a wall up in front of me. 
    And I have to lean back to see the top. 
    And really the only constraint in that classroom is that it's terrifying. 
    It is the most terrifying experience I've ever had.\\
    \textbf{First author:} All eyes on you. \\
    Exactly. 
    I'm waiting for a pit to open and violence to come out. 
    It's a really intense room to teach in.} (P08, instructor)
\end{quote}

The instructor teaches in a tiered lecture hall with hundreds of students. 
The \textbf{performative nature} associated with live coding created anxiety.
With experience, the stress of live coding is reduced according to teaching assistant P06 and instructors P08 and P11.

In reacting to asking student to code in front of the class, a teaching assistant noted the punitive nature of editors over classical whiteboard coding.
Unlike the free form acceptance of a whiteboard, the editor punishes: 

\begin{quote}
    \textit{If there's a spelling mistake [when white boarding], or if [a student coding on a whiteboard in front of the class] miss[es] a comma or something, no one cares... 
    You do that on a computer, then \textbf{it'll scream at you}, and then there will be the red squiggly.} (P02, teaching assistant)
\end{quote}

Annotations from the editor call out mistakes, which can embarrass a student who is coding in front of the class.
The intimidating nature of live coding, exacerbated by a large audience, prevents P08 from asking his students to code in front of their peers.

\begin{quote}
    \textit{I worry about [asking students to code in front of the class] in my 100 person classroom, it's intimidating for me sometimes and so I could see that would be really \textbf{intimidating for students}.} (P08, instructor)
\end{quote}

The negative emotions associated with live coding sometimes make it difficult a teacher and student to code in front of the class. 

\subsection{Benefits of Live Coding}

Our analysis also uncovered perceived positive aspects of the practice.
Findings corroborate much prior work on student's perceptions on the benefits of live coding \cite{DBLP:conf/kolicalling/RajPHH18}. 

\paragraph{Modelling programming strategies}

Unlike presenting complete, static code, making mistakes while live coding helps students to learn.
The instructor making mistakes can help address issues of imposter syndrome, demonstrate productive debugging strategies, and model ways of identifying and pursuing promising paths to take after introducing an error.
These mistakes also provide potentially unplanned opportunities for student engagement.

The power and pedagogical productive for instructors making errors while live coding was commented on by a student who said

\begin{quote}
    \textit{... when he makes mistakes, it's like that's a good example of what not to do. 
    And if he makes the code beforehand then you're not really gonna see that.}
    (P09, student)
    \end{quote}

Explicit programming strategies help with problem solving~\cite{latoza2020explicit}.
To learn \textbf{programming strategies}, students can observe how instructor's recover from mistakes.
At the same time, these mistakes can serve as teaching opportunities for instructors to talk about the at-times messy practice of programming.

\paragraph{Potential for active learning}

A tenant of live coding is the demonstration of the incremental process of programming~\cite{ali2005effective}.
By creating a treasure-hunt for mistakes while live coding, some instructors harness students' attention in the learning activity. 
In some cases, live coding affords opportunities for instructors to \textbf{react} to students and to \textbf{change on the fly}, leading to reports of student engagement.

\begin{quote}
    \textit{I'm doing some example, then it's easier to change stuff on the fly and then surprise students.} (P02, teaching assistant)
\end{quote}

Asking students to make \textbf{predictions} and incorporating their suggestions is a pedagogical practice that can accompany live coding and was mentioned by several instructors (P08, instructor, P12, teaching assistant, and P14, teaching assistant). 

\begin{quote}
    \textit{So when it goes really well is when you can get students to participate. 
    So it's like, you know, you're doing some code and it's like, okay, what's the next step and you get feedback and you can be like, oh, well, that's a great suggestion. 
    Let's try that and maybe it doesn't work out and then you can say, okay, so how do we adjust?} (P08, instructor)
\end{quote}

Live coding allows for flexibility to introduce areas that the instructor has not yet explored.
Unlike slide-based lectures and static code examples, where all material is decided upon ahead of time, our participants felt that live coding supports student-centered learning by incorporating suggestions from the student and that might not have been considered by the instructor.
During live coding, the instructor can adjust the difficulty by changing the content of examples and the amount of skeleton code provided.
Benefits in student engagement, specifically in opportunities for student predictions, corroborate prior work~\cite{watkins2024comparing}.
In this way, live coding enables the instructor to make changes in real time and can thus be more responsive to their class and better meet the students at their current level of ability.
By \textbf{adjusting the difficulty}, the instructor can increase engagement.

\section{Discussion}

This analysis of instructor and student experiences with live coding in large introductory computer science classrooms revealed insights into barriers associated with live coding as well as benefits of the pedagogical approach.
We interpreted our findings, delving into possible considerations to overcome problems in live coding and ways to support effective live coding.
Here we propose practical and technological implications for consideration when building live coding tools. 

\begin{table*}
    \centering
    \begin{tabular}{p{0.3\linewidth} p{0.3\linewidth} p{0.3\linewidth}}

    \toprule
    \textsc{Benefits} & \textsc{Barriers} & \textsc{Implications}\\
    \midrule
    Modelling programming strategies & Teaching environment: inadequate & Proper physical classrooms\\
    Potential for active learning & resources and setup  & Personal computers optional\\
     & Live coding is scary & Directing attention \\
     & & Teacher growth \\
     & & Community and psychological safety \\
    \bottomrule
    \end{tabular}
    \caption{\textbf{Implications for live coding.} Benefits, barriers, and implications of live coding.}
    \label{tab:livecodingimpl}
\end{table*}

\paragraph{Teacher growth}

Our findings indicate that although instructors and students reported educational benefits, live coding is a daunting task. 
New instructors and teaching assistants might find \textbf{support and advice from senior faculty} to gain confidence in live coding. 
In addition to reducing discomfort, experienced instructors could share practices and techniques that support the reported educational benefits of live coding.

Live coding might require new instructors (or seasoned instructors teaching an unfamiliar topic) to abandon the notion that they are experts. 
Effective live coding involves making mistakes in front of the classroom and incorporating student suggestions.
Debugging faulty code under time pressure and the watchful eyes of a hundred students adds significant difficulty to an already cognitively taxing task.
This often requires a \textbf{mindset change for instructors} accustomed to smooth and structured slides-only class delivery methods.
In other words, the instructor must get used to not being an infallible expert, but a guide in the collaborative learning process for the entire classroom.
In the design and selection of classrooms for introductory computer science courses, our findings suggest identifying \textbf{room layouts that elicit comfort and confidence} as opposed to anxiety amongst instructors, and students who code in front of their peers.

\paragraph{Proper physical classrooms}

The podium should have ample space for instructors to organize their materials and the necessary equipment so that the instructor can live code comfortably. 
The projection screens should be large enough (or there should be multiple ones in different locations) so that all students can see code on the projector, regardless of their position in the room.
There should be space---ideally a secondary screen---for the instructor to keep the correct source code (a ``cheat sheet'') as a reference.

If the lesson is designed so that students are asked to code in front of their peers, there should be outlets in the classroom or the session should take place in a computer lab. 
Similarly, necessary material (e.g., whiteboard, paper, and pens) should be available for alternate forms of participation. 

\paragraph{Community and psychological safety}

Students expressed different levels of comfort sharing their code with classmates.
To encourage students to code in front of their peers, tools should be designed with \textbf{psychological safety} in mind.
Instead of editors that punish with red squiggly lines that highlight a missing comma, tools should have modes that encourage risk taking, and trial and error over a perfectly correct piece of code.
Tools should elicit feelings of pride over embarrassment among students by supporting the communication of a programming strategy over highlighting a syntactic mistakes of the resulting program.
To accomplish this, tool could support a pseudo code, white boarding mode in addition to traditional editors.
Beyond tooling and arguably more important than a technical solution, it is important to create a classroom culture and wider department and university setting where students feel comfortable making mistakes and reaching out to their peers for help, in instructor P11's words, an \textit{``open atmosphere''}.
If students can be vulnerable with their classmates, peer instruction and student-led live coding activities are more feasible, thus leading to a more active form of learning.

\paragraph{Personal computers optional}

In terms of technology, during lessons where the instructor live codes, students should be able to \textbf{opt in or out of using their own personal laptops}. 
Some students and instructors expressed that computers sometimes acted as a distractor. 
Some preferred to listen only or take notes via pen and paper; student P03 reports: \textit{``I've always taken hand notes''.}
The students' sentiments bolsters prior work that found that laptop multitasking had a negative impact on learning for students (both the multitasker and their neighbors) in classroom settings~\cite{sana2013laptop}.

Students should be able to follow the lecture in a way that best serves their personal learning preferences. 
This guideline conflicts with prior work on live coding tools because it challenges the assumption that student participation should occur through students' personal laptops.
Not requiring personal computers helps to facilitate student attention to the pedagogically beneficial parts of instructor-led live coding (e.g., modelling mistakes).

\paragraph{Directing attention.}

Tools should support \textbf{directing student attention} to areas of the code of importance.
Teaching assistant P14 reported that her students struggle to identify what they should be looking at and paying attention to. 
Live coding tools should direct student attention to which part of the code is being discussed so that students are synchronizing the instructors words with the intended piece of code.
If students are looking at the right portion of the code, they can better observe the process of creating programs and debugging.

For example, live coding tools could support giving instructor's control of which areas of their screen are projected so that only a portion of a window is projected.
Increased control over screen sharing will also help the projector-to-screen ratio be manageable from the perspective of the instructor and readable from the perspective of the student in a large lecture hall. 
Highlighting or zooming focused code, or dimming unrelated code, are possible mechanisms to concentrate attention. 

\subsection{Limitations}

Despite presenting compelling dimensions of live coding, our study is limited in terms of practical outcomes and actual tooling support.
Due to the formative nature of our study and small participant count, our sample is not an exhaustive representation of the broader group of computer science students, teaching assistants, and instructors. 
Further, half of P04's data was lost due to technical problems, thereby reducing the corpus of data that was analyzed.  
Identity attributes, such as gender, race, and socioeconomic factors, among others, contribute to the diversity---or lack thereof---of the population.
Although the instructors' and teaching assistants' number of years of experience would provide context, we decided against reporting such information to protect the anonymity of participants.
Overall, the small sample limits the generalizability of our findings.

Since the work was conducted in a U.S.\ public university, findings might not translate to other educational settings or geographic contexts.
Furthermore, the set of barriers, benefits, and proposed design guidelines are not exhaustive.
Finally, we have not yet validated the design guidelines in a real live-coding tool.
Collectively, more work is required to gain further insight into live coding as a pedagogical practice.

\subsection{Future Work}

The list of barriers, facilitators and implications is not exhaustive.
Future work could expand these findings by running future studies with groups of participants with differing backgrounds from the participants in our study.
Furthermore, we plan to validate the proposed design implications.

Participants noted that there might be a connection between our work and tools for teaching writing.
Additional research could investigate the relationship between live writing and live coding tools. 
Outside of the introductory to programming classroom context examined in this paper, participants highlighted the applicability of live coding tools to support knowledge transfer in hobbyist communities and informal learning environments, like robotics clubs.
Continued work should explore live coding in other settings.
Future work should consider new audience engagement hardware for live coding due to drawbacks in personal laptops.

\section{Conclusion}

We have presented a study involving instructors, teaching assistants, and students to uncover barriers and benefits of live coding in computer science classrooms.
We then use these findings to propose design guidelines for live coding support tools.  
Live coding can be scary for the instructor and the students.
Yet, pedagogical benefits exist to live coding such as modeling debugging strategies and encouraging student involvement. 
We recommend tools remove or reduce barriers and support positive aspects of live coding, and present additional design guidance for attributes, tasks, and features that live coding tools could support.

\subsubsection*{Acknowledgments}
    We thank the anonymous reviewers for their feedback on this paper.
    Any opinions, findings, and conclusions or recommendations expressed here are those of the authors and do not necessarily reflect the views of the funding agency.
    We thank participants for their time and candor and for sharing their live coding experiences, both negative and positive.
    We appreciate Saransh Grover's feedback during data analysis, Clemens Klokmose's suggestions on interview techniques, and Joel Chan's thoughts during early scoping of the work.     

\bibliographystyle{splncs04}
\bibliography{live-coding}

\end{document}